\documentclass[preprint,prd,aps,nofootinbib]{revtex4}
\usepackage{amsmath}
\usepackage{graphicx}
\usepackage{diagbox}
\usepackage{subcaption} 
\usepackage{tikz}
\usepackage{slashed}
\usepackage{float}
\usepackage{color}
\begin{document}
	\title{Relativistic Bethe-Salpeter study of OZI-rule allowed strong decays: determination of total width of $h_{c}(2P)$ and $^{3}P_{0}$ model parameter}
	\author{Yi-Yi Rui$^{1,2,3,4}$, Tianhong Wang$^5$, Zhi-Hui Wang$^{6,7}$, Tai-Fu Feng$^{1,2,3,4}$, 
		Guo-Li Wang$^{1,3,4}$\footnote{corresponding author}}
	\affiliation{$^1$ Department of Physics, Hebei University, Baoding 071002, China\\
         $^2$Department of Physics, Guangxi University, Nanning, 530004, China\\
		$^3$ Hebei Key Laboratory of High-precision Computation and Application of Quantum Field Theory, Baoding 071002, China\\
		$^4$ Hebei Research Center of the Basic Discipline for Computational Physics, Baoding 071002, China\\
		$^5$ School of Physics, Harbin Institute of Technology, Harbin 150001, China\\
		$^6$Key Laboratory of Physics and Photoelectric Information Functional Materials, North Minzu University,
		Yinchuan 750021, China\\
		$^7$School of Electrical and Information Engineering, North Minzu University, Yinchuan 750021, China}
	\begin{abstract}
		Strong decays allowed by the Okubo-Zweig-Iizuka rule play a decisive role in determining the properties of particles. The widely used $^{3}P_{0}$ model is non-relativistic and contains unknown adjustable parameter $\gamma$ that need to be determined experimentally. {Based on the Bethe-Salpeter equation, we have derived a relativistic calculation formula in which the effective strength related to the $^{3}P_{0}$ parameter $\gamma$ is not introduced as an independent fitting parameter, but is consistently determined by the interaction kernel of the Bethe–Salpeter equation.} Using this method, we present the total width of $h_{c}(2P)$ and {allows a theoretical determination of the parameter $\gamma$ in the $^{3}P_{0}$ model within the present framework.}
	\end{abstract}

	\maketitle
	
	\section{Introduction}
	
	The partial widths for open-flavor strong decays of particles, i.e., Okubo-Zweig-Iizuka (OZI) rule-allowed strong decays \cite{O,Z,I}, are generally much larger than those for OZI rule-forbidden strong decays, electromagnetic decays, or weak decays. Consequently, these decays play a decisive role in determining particle properties, such as estimating a particle's total width (lifetime).
	
	Theoretically, there are multiple methods for calculating OZI rule-allowed strong decays, such as: heavy quark symmetry \cite{hill}, effective Lagrangian based on the chiral quark model \cite{pierro,zhaoq1,zhaoq2}, the method employing reduction formulae combined with low-energy theorems and Partial Conservation of Axial Current \cite{wgla,wglb}, effective Lagrangian approach preserving heavy quark spin-flavor and light quark chiral transformations \cite{Colangelo,Fazio}, and heavy quark effective theory with chiral symmetry breaking corrections \cite{wzg}, etc. However, all these methods have limitations in application and primarily address the decay of one heavy meson into another heavy meson plus a light one.
	
	The most widely used and unrestricted approach is the $^3P_0$ model, also known as the quark-pair creation model. This model was initially proposed by Micu \cite{micu} and subsequently extended by Orsay group, who pioneered its application to hadronic decays \cite{Yaouanc1,Yaouanc2,Yaouanc3}. The model contains a parameter $\gamma$ (or $g=2m_q\gamma$), which reflects the strength of light quark-pair creation (with $0^{++}$ quantum number in a $^3P_0$ state, therefore called the $^3P_0$ model) from the vacuum and needs to be determined by fitting experimental data. It has been found that $\gamma$ exhibits a certain degree of light quark flavor independence and typically takes values between 0.3 and 0.5. Thus, the $^3P_0$ model possesses reasonable predictive power and {is widely adopted \cite{Godfrey3,Barnes2,Barnes3,WZH26,gamaqu0.35,Godfrey4,spinningpairs,gamastate,charmoniumproduct,
charmonia,ds0ds1,bbaryon,sdcharmonia,3p0revisited,charmedmeson,bottomonia,k2600,charmoniumstate,
omega2012,roumeson,1fwave,3s2d,kaons,psigrrr,strangeonium,y4500}}. Furthermore, it has several variants and has been extended numerous times, for example, the Cornell group postulated a new microscopic model for the QCD mechanism underlying strong decays by assuming that light quark pairs are produced from the linear confining interaction, while also accounting for coupled-channels effects \cite{Eichten0,Eichten1,Eichten2}, Kokoski \textit{et al.} modulated the spatial dependence of the pair production amplitude to simulate a gluonic flux tube \cite{Isgur1,Godfrey2}, Ackleh \textit{et al.} investigated quark pair production mechanisms mediated by both the scalar confining interaction and one-gluon exchange \cite{Barnes1}, Fu \textit{et al.} extended the nonrelativistic $^3P_0$ model by incorporating relativistic corrections \cite{fhf13}, and Segovia \textit{et al.} generalized the $\gamma$ parameter of the $^3P_0$ model to a scale-dependent form \cite{segovia}, etc.
	
	However, the $^3P_0$ model and its variants still have limitations. First, the original theory and most of its extensions are non-relativistic, while the final mesons contain light quarks, implying that relativistic corrections cannot be neglected. Second, empirical evidence shows that the $\gamma$ values for the creation of $u$(or $d$) quark pairs and $s$ quark pairs from the vacuum are distinct. Finally, and most significantly, the parameter $\gamma$ must be determined experimentally, resulting in limited predictive power for the model. 
{Therefore, this paper presents a method for calculating OZI rule-allowed strong decays based on the Bethe--Salpeter equation. In this approach, the effective decay strength is not introduced as an independent phenomenological parameter analogous to $\gamma$, but is intrinsically connected to the same interaction kernel that also determines hadron spectra. As a result, the $^{3}P_{0}$ coupling can be related to the underlying quark--antiquark dynamics, enhancing the predictive power of the formalism.}
Furthermore, the parameters of the potential model can also be determined by calculating the mass spectrum (or other decay processes), endowing the method with strong predictive power. Moreover, this approach constitutes a relativistic theoretical framework.
	
	\section{An approach to OZI rule-allowed strong decays via the Bethe-Salpeter equation}
	
	The Bethe-Salpeter (BS) equation is a relativistic dynamical equation describing bound states composed of two particles \cite{bs1951}. In momentum space, the BS equation describing a meson composed of quark 1 and antiquark 2 is:
	\begin{eqnarray}\label{eq1}
		\chi_{_P}(q)=i S_{1}(p_{1}) \int\frac{d^{4}k}{(2\pi)^{4}}I(P;q,k)\chi_{_{P}}(k)S_{2}(-p_{2}),
	\end{eqnarray}
	where $p_1$ and $p_2$ are the momenta of quark 1 and antiquark 2, $P$ and $q$ are the meson's total momentum and internal relative momentum. They satisfy the relations
	$p_{1}=\frac{m_{1}}{m_{1}+m_{2}}P+q$, $p_{2}=\frac{m_{2}}{m_{1}+m_{2}}P-q$, where $m_1$ and $m_2$ are the masses of quarks.
	$\chi_{_P}(q)$ is the meson's relativistic wave function. $S_{1}(p_{1})$ and $S_{2}(-p_{2})$ are the propagators of the quarks. $I(P;q,k)$ is the interaction kernel between the quark and antiquark. 
	
	\begin{figure}[!htbp]
		\begin{center}\label{BS1}
			\begin{minipage}[c]{1\textwidth}
				\includegraphics[width=2.5in]{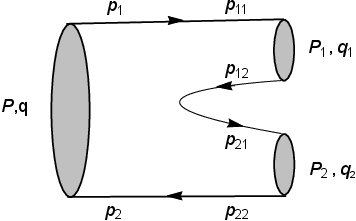}
				\includegraphics[width=2.5in]{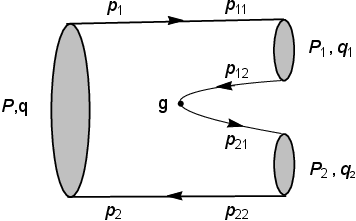}
			\end{minipage}%
			\caption[]{Feynman diagrams of OZI rule-allowed strong decays within the BS approach (left) and the $^{3}P_{0}$ model (right).}
		\end{center}
	\end{figure}

	We noted that in Eq. (\ref{eq1}) the BS wave function $\chi_{P}(q)$ contains two external legs, namely the propagators $S_1$ and $S_2$. This feature conveniently allows us to express the transition amplitude using the BS wave function. For example, the Feynman diagram in Fig. 1 depicts an OZI rule-allowed strong decay of a doubly heavy meson into two heavy-light mesons (where we have labeled the meson momentum and its relative momentum, as well as the quark momenta. Quark masses, labeled analogously to their momenta, are not repeated here). According to the Mandelstam mechanism \cite{mandelstam}, the transition $S$-matrix element for this decay can be expressed as
	\begin{eqnarray}\label{amplitude}
		\begin{split}
			\langle P_{1}P_{2}|S|P\rangle_{BS}=C_{f}\int\frac{d^{4}q}{(2\pi)^{4}}\frac{d^{4}q_{1}}{(2\pi)^{4}}\frac{d^{4}q_{2}}{(2\pi)^{4}}
			Tr[\chi_{_{P}}(q)S^{-1}_{2}(-p_{2})(2\pi)^{4}\delta^{4}(p_{2}-p_{22})\overline{\chi}_{_{P_{2}}}(q_{2})
			\\ \times S^{-1}_{1}(p_{21})(2\pi)^{4}\delta^{4}(p_{21}+p_{12})\overline{\chi}_{_{P_{1}}}(q_{1})S^{-1}_{1}(p_{1})(2\pi)^{4}\delta^{4}(p_{1}-p_{11})],
		\end{split}
	\end{eqnarray}
	where $C_{f}=\frac{1}{\sqrt{3}}$ is the color factor, and $\overline{\chi}_{_{P_{i}}}(q_{i})=\gamma^{0}\chi^{\dagger}_{P_{i}}(q_{i})\gamma^{0}$,   $i=1,2$ for final mesons 1 and 2.
	Here, the Feynman rules are similar to the standard ones. First, write the wave function $\chi_{_{P}}(q)$ for the initial meson, then write the wave function for the final meson 2 backwards along the fermion line. We note that the external leg of antiquark 2 is shared by both the initial meson and final meson 2, see diagrams in Fig. 1, thus being occupied twice. Therefore, on one hand, we must compensate with an inverse propagator $S^{-1}_{2}(-p_{2})$; on the other hand, we also need to add a delta function $(2\pi)^{4}\delta^{4}(p_{2}-p_{22})$ to ensure energy-momentum conservation, and so on. Second, due to the formation of a closed loop, there is an overall trace $Tr[...]$. Finally, also due to the closed loop,  we need to integrate out the internal momenta $q$, $q_1$, and $q_2$.
	
	As evidenced in the derivation of transition amplitude Eq. (\ref{amplitude}), our approach does not adopt the $^3P_0$ model hypothesis, namely, we do not assume that light quark-antiquark pairs are created from the vacuum. Consequently, there is no need to introduce the strength factor $g$ or $\gamma$ for light quark pair creation. From the BS Eq. (\ref{eq1}), it is evident that a constituent quark and an antiquark are bound as a meson through the interaction kernel $I$ between quarks. Therefore, OZI rule-allowed strong decay processes also originate from this interaction kernel $I$.
	
	Applying the relations between the meson's momentum and its internal momentum, and exploiting the delta function to integrate out $q_1$ and $q_2$, Eq. (\ref{amplitude}) is transformed into:
	\begin{eqnarray}\label{amplitude2}
		\begin{split}
			&\langle P_{1}P_{2}|S|P\rangle_{BS}\equiv(2\pi)^{4}\delta^{4}(P-P_{1}-P_{2})\mathcal{M}_{BS}\\
			&=(2\pi)^{4}\delta^{4}(P-P_{1}-P_{2}) \frac{1}{\sqrt{3}} \int\frac{d^{4}q}{(2\pi)^{4}}
			Tr[\chi_{_{P}}(q)S^{-1}_{2}(-p_{2})\overline{\chi}_{_{P_{2}}}(q_{2})S^{-1}_{1}(p_{21})\overline{\chi}_{_{P_{1}}}(q_{1})S^{-1}_{1}(p_{1})],
		\end{split}
	\end{eqnarray}
	where $q_{1} =q+(\frac{m_{1}}{m_{1}+m_{2}} P-\frac{m_{11}}{m_{11}+m_{12}}P_{1})$ and $q_{2} =q+(-\frac{m_{2}}{m_{1}+m_{2}} P+\frac{m_{22}}{m_{21}+m_{22}}P_{2})$.
	
	To proceed, it is necessary to solve the BS equation; however, rather than solving the challenging BS equation, we employ Salpeter's approach \cite{salpeter} by applying the instantaneous approximation to the BS equation and then solving it.
	The instantaneous approximation is well-suited for heavy mesons, and assumes that the propagation time of interactions between quarks within the meson is negligible, meaning the kernel depends only on spatial momenta:
	$I(P;q,k)\simeq I(q_{\perp},k_{\perp})= I(q_{\perp}-k_{\perp})$, where we have defined ${q}_{_{\perp}}={q}-\frac{P}{M}q_{_{P}}$ with $q_{_{P}}\equiv\frac{P\cdot q}{M}$. In the meson's center-of-mass system, $P = (M, 0)$, $q_{_P} = q_{_0}$, and ${q}_{_{\perp}}=(0,\vec{q})$. 
	
	After adopting the instantaneous approximation, we have
	\begin{eqnarray}\label{right}
		i \int\frac{d^{4}k}{(2\pi)^{4}}I(P;q,k)\chi_{_{P}}(k)
		=\int\frac{d^{3}k_{_{\perp}}}{(2\pi)^{3}}I(q_{_{\perp}}-k_{_{\perp}})\varphi_{_{P}}(k_{_{\perp}})\equiv \eta_{_{P}}(q_{_{\perp}}),
	\end{eqnarray}
	where $\varphi_{_P}(k_{_{\perp}})\equiv i\int\frac{dk_{_P}}{2\pi}\chi_{_P}(k)$ is the Salpeter wave function.
	Since we are providing a relativistic wave function, a non-relativistic interaction kernel must be adopted to avoid double-counting. Therefore, we choose the modified Cornell potential \cite{Ding,kernel}:
	\begin{eqnarray}\label{potential2}
		I(\vec{q}) = -(\frac{\lambda}{\alpha}+V_{0})\delta^{3}(\vec{q})+\frac{\lambda}{\pi^{2}}\frac{1}{(\vec{q}^{2}+\alpha^{2})^{2}}
		-\gamma_{0}\otimes\gamma^{0}\frac{2}{3\pi^{2}}\frac{\alpha_{s}(\vec{q})}{(\vec{q}^{2}+\alpha^{2})},
	\end{eqnarray}
	{where $V_0$ is a free parameter to fit mesons mass}, $\lambda$ is the string tension, $\alpha$ is a small quantity introduced to avoid infrared divergence and account for screening effects, and
	$\alpha_{s}(\vec{q})=\frac{12\pi}{27}\frac{1}{log(e+\frac{\vec{q}^{2}}{\Lambda_{QCD}^{2}})}$ is running coupling constant, with $\Lambda_{QCD}$ is the QCD scale and $e=2.7183$.
	
	With Eq. (\ref{right}), the BS equation in Eq. (\ref{eq1}) reduces to
	\begin{eqnarray}\label{halfBS}
		\chi_{_{P}}(q)=S_{1}(p_{1})\eta_{_{P}}(q_{_{\perp}})S_{2}(-p_{2}).
	\end{eqnarray}
	The propagators can be written as
	\begin{eqnarray}
		\begin{split}\label{propagator}		-iS_{1}(p_{1})=\frac{\Lambda^{+}_{1}}{p_{_{1P}}-\omega_{1}+i\varepsilon}+\frac{\Lambda^{-}_{1}}{p_{_{1P}}+\omega_{1}-i\varepsilon},~ iS_{2}(-p_{2})=\frac{\Lambda^{+}_{2}}{p_{_{2P}}-\omega_{2}+i\varepsilon}+\frac{\Lambda^{-}_{2}}{p_{_{2P}}+\omega_{2}-i\varepsilon},
	\end{split}\end{eqnarray}
	where $\omega_{i}=\sqrt{m^{2}_{i}-p^{2}_{i\perp}}$, and the projection operator
	$ \Lambda^{\pm}_{i}=\frac{1}{2\omega_{i}}[\frac{\slashed{P}}{M}\omega_{i}\pm(\slashed{p}_{i\perp}+(-1)^{(i+1)}m_{i})]
	$.

	If define
	$\varphi^{\pm\pm}\equiv\Lambda^{\pm}_{1}
	\frac{\not{P}}{M}\varphi\frac{\not{P}}{M} \Lambda^{\pm}_{2}$,
	then we have
	$\varphi=\varphi^{++}+\varphi^{+-}+\varphi^{-+}+\varphi^{--}$.
	Integrating both sides of Eq. (\ref{halfBS}) over $q_{_P}$, applying Eq. (\ref{propagator}) along with the residue theorem, and the relations of the projection operators:
	$\Lambda^{+}_{i}+\Lambda^{-}_{i}=\frac{\slashed{P}}{M},~
	\Lambda^{\pm}_{i}\frac{\slashed{P}}{M}\Lambda^{\pm}_{i}=\Lambda^{\pm}_{i},~
	\Lambda^{\pm}_{i}\frac{\slashed{P}}{M}\Lambda^{\mp}_{i}=0$, we obtain the Salpeter equation:
	\begin{eqnarray}\label{sal2}
		\begin{split}
			&
			(M-\omega_{1}-\omega_{2})\varphi^{++}_{_P}(q_{_\perp})=
			\Lambda^{+}_{1}
			\eta_{_P}(q_{_\perp})\Lambda^{+}_{2},\\
			&
			(M+\omega_{1}+\omega_{2})\varphi^{--}_{_P}(q_{_\perp})=
			-\Lambda^{-}_{1}
			\eta_{_P}(q_{_\perp})\Lambda^{-}_{2},\\
			&
			\varphi^{+-}_{_P}(q_{_\perp})=\varphi^{-+}_{_P}(q_{_\perp})=0,
		\end{split}
	\end{eqnarray}
	where $\varphi^{++}$ and $\varphi^{--}$ are the positive-energy and negative-energy wave functions, respectively. For bound states, the relation $M-\omega_{1}-\omega_{2}\ll M+\omega_{1}+\omega_{2}$ generally holds, which implies $\varphi^{++}\gg \varphi^{--}$. Therefore, in the subsequent analysis, we will neglect the contribution from the $\varphi^{--}$. 
	
	Substituting the BS equation Eq. (\ref{halfBS}) into the formula Eq. (\ref{amplitude2}), we obtain the expression for the transition amplitude:
	\begin{eqnarray}
		\begin{split}
			&\mathcal{M}_{BS}=\frac{1}{\sqrt{3}}\int\frac{d^{4}q}{(2\pi)^{4}}Tr[S_{1}(p_{1})\eta_{_{P}}(q_{_\perp})S_{2}(-p_{22})\overline{\eta}_{_{P_{2}}}(q_{_{2\perp}})S_{2}(-p_{12})\overline{\eta}_{_{P_{1}}}(q_{_{1\perp}})]\,\\
			&= \frac{-i}{\sqrt{3}}\int\frac{d^{4}q}{(2\pi)^{4}}Tr\Bigg[\frac{\Lambda^{+}_{1}}{p_{_{1P}}-\omega_{1}+i\varepsilon}\eta_{_{P}}\frac{\Lambda^{+}_{2}}{p_{_{2P}}-\omega_{2}+i\varepsilon}\overline{\eta}_{_{P_{2}}}
			\Big(\frac{\Lambda^{+}_{12}}{p_{_{12P}}-\omega_{12}+i\varepsilon}-\frac{\Lambda^{+}_{21}}{p_{_{21P}}-\omega_{21}+i\varepsilon}\Big)\overline{\eta}_{_{P_{1}}}\Bigg].
		\end{split}
	\end{eqnarray}
	Here, due to the dominance of the $\varphi^{++}$, we have neglected the contributions from the negative-energy projection operator in the first two propagators; in the third one, we have further used the relations $\Lambda^{-}_{12}=\Lambda^{+}_{21}$ and $p_{_{12P}}+\omega_{12}-i\varepsilon=-(p_{_{21P}}-\omega_{21}+i\varepsilon)$.
	
	After performing the integration over $q_{_P}$ on the right-hand side of the above equation and using $\omega_{1}=\omega_{11}$ and $\omega_{2}=\omega_{22}$, we obtain:
	\begin{eqnarray}
		\begin{split}\label{amplitude3}
			\mathcal{M}_{BS}= \frac{-1}{\sqrt{3}}\int\frac{d^{3}q_{_{\perp}}}{(2\pi)^{3}}Tr\Bigg[\frac{\Lambda^{+}_{1}\eta_{_{P}}
				\Lambda^{+}_{2}}{M-\omega_{1}-\omega_{2}}\overline{\eta}_{_{P_{2}}}\Big(\frac{\Lambda^{+}_{12}}{P_{_{1P}}-\omega_{11}-\omega_{12}}
			-\frac{\Lambda^{+}_{21}}{P_{_{2P}}-\omega_{22}-\omega_{21}}\Big)\overline{\eta}_{_{P_{1}}}  \Bigg]\\
			=
			\frac{1}{\sqrt{3}}\int\frac{d^{3}q_{_{\perp}}}{(2\pi)^{3}}Tr\Bigg[\varphi^{++}_{_P}\frac{\slashed{P}}{M}
			\frac{\Lambda^{+}_{2}~\overline{\eta}_{_{P_{2}}}\Lambda^{+}_{21}}{P_{_{2P}}-\omega_{22}-\omega_{21}}\overline{\eta}_{_{P_{1}}}
			-\frac{\slashed{P}}{M}\varphi^{++}_{_P} \overline{\eta}_{_{P_{2}}} \frac{\Lambda^{+}_{12}~\overline{\eta}_{_{P_{1}}} \Lambda^{+}_{1}}{P_{_{1P}}-\omega_{11}-\omega_{12}}   \Bigg]\\
			=
			\frac{1}{\sqrt{3}}\int\frac{d^{3}q_{_{\perp}}}{(2\pi)^{3}}Tr\Bigg[\varphi^{++}_{_P}(q_{_{\perp}})\frac{\slashed{P}}{M}
			\overline{\varphi^{++}_{_{P_{2}}}}(q_{_{2\perp}})
			\overline{\eta}_{_{P_{1}}}(q_{_{1\perp}})
			-\frac{\slashed{P}}{M}\varphi^{++}_{_P}(q_{_{\perp}}) \overline{\eta}_{_{P_{2}}}(q_{_{2\perp}})
			\overline{\varphi^{++}_{_{P_{1}}}}(q_{_{1\perp}}) \Bigg],
		\end{split}
	\end{eqnarray}
	where $q_{_{1\perp}} =q_{_{\perp}}-\frac{m_{11}}{m_{11}+m_{12}}P_{_{1\perp}}$ and $q_{_{2\perp}} =q_{_{\perp}}+\frac{m_{22}}{m_{21}+m_{22}}P_{_{2\perp}}$.
	Thus, we have established a relativistic method for calculating OZI rule-allowed strong decays based on the BS equation and the screened Cornell potential. 
{This approach does not treat the strength of light quark pair production as an independent fitting parameter $\gamma$. Instead, the decay amplitude depends on the vertex function $\eta_{P}(q_{\perp})$, which is fully determined by the interaction kernel of the BS equation. As a result, the effective $^{3}P_{0}$ coupling is not arbitrary, but originates from the same dynamics that govern the hadron spectrum. This allows us to estimate the parameter $\gamma$ in the $^{3}P_{0}$ model, since the same interaction kernel is employed for both the decay and the spectrum.}

	\section{$^{3}P_{0}$ model}
	
	The $^3P_0$ model relies on the assumption that a light quark pair is produced from the vacuum with a strength $g = 2m_q\gamma$, as depicted in the right diagram in Fig. 1. Unlike left one in Fig. 1, due to the presence of the $g$ vertex, the light quarks do not share a common propagator (and consequently, the color factor $C_f$ does not appear); instead, they have their respective propagators. The transition amplitude is then given by \cite{fhf13}:
	\begin{eqnarray}\label{3P0am}
		\begin{split}
			\mathcal{M}_{^3P_0}=g\int\frac{d^{4}q}{(2\pi)^{4}}
			Tr[\chi_{_{P}}(q)S^{-1}_{2}(-p_{2})\overline{\chi}_{_{P_{2}}}(q_{2})\overline{\chi}_{_{P_{1}}}(q_{1})S^{-1}_{1}(p_{1})]\\
			=g\int\frac{d^{3}q_{_\perp}}{(2\pi)^{3}}
			Tr[\frac{\slashed{P}}{M}\varphi^{++}_{_P}(q_{_{\bot}})
			\frac{\slashed{P}}{M}\overline{\varphi^{++}_{_{P_{2}}}}(q_{_{2\bot}})\overline{\varphi^{++}_{_{P_{1}}}}(q_{_{1\bot}})](1-\frac{M-\omega_{1}-\omega_{2}}{2\omega_{12}}).
		\end{split}
	\end{eqnarray}
	
	Comparing Eq. (\ref{3P0am}) with Eq. (\ref{amplitude3}), we note that the main difference between the $^{3}P_{0}$ model and the BS method is that the amplitude in the $^{3}P_{0}$ method involves the overlap integral of three meson wave functions, and the production of light quark pairs is determined by the strength factor $g$. In contrast, in the BS method, one of the final meson wave functions is replaced by the result after being acted upon by the interaction potential. This indicates that the origin of light quark pair production lies in the interaction between quarks.
	
	\section{OZI allowed strong decay $h_c(2P)\to D\bar{D}^*$ ($D^*\bar{D}$)}
	To verify the correctness of the theory, we employed the two methods above to calculate the OZI-allowed strong decay of $h_c(2P)$. This process possesses inherent research value, as the charmonium $1^{+-}$ state $h_c(2P)$ has not yet been experimentally observed. Theoretically, its mass lies above the $D\bar{D}^*$ ($D^*\bar{D}$) threshold; consequently, its dominant decay modes are strong decays $h_c(2P)\to D\bar{D}^*$ ($D^*\bar{D}$), and it is expected to have a relatively wide width. Thus, studying its strong decays is of significant importance for understanding this particle.
	
	To compute the transition amplitude, we need to solve the Salpeter equation Eq. (\ref{sal2}) to obtain the corresponding wave functions for $1^{+-}$, $0^{-}$ and $1^{-}$ mesons. This task has been completed in our previous papers \cite{kernel,1P1,wave1-}; thus, only the representations of the wave functions are provided here. 
	$h_c(2P)$ has four OZI-allowed decay channels: $h_c(2P) \to D^0 \bar{D}^{*0}$, $h_c(2P) \to D^{*0} \bar{D}^0$, $h_c(2P) \to D^+ D^{*-}$, and $h_c(2P) \to D^{*+} D^-$. We take $h_c(2P) \to D^0 \bar{D}^{*0}$ as an example to present the wave functions. That is, the final particle 1 is $D^0$, and particle 2 is $\bar{D}^{*0}$.
	For the $1^{+-}$ meson $h_c(2P)$, its relativistic wave function is \cite{1P1}
	\begin{equation}
		\varphi^{{1^{+-}}}_{_P}(q_{_{\bot}})=\epsilon\cdot q_{_{\bot}} \left(a_{1}+a_{2}\frac{\slashed{P}}{M}
		+a_{3}\frac{\slashed{P}\slashed{q}_{_{\bot}}}{M^2}\right)\gamma_{5},
	\end{equation}
	where, $P$ and $M$ are the momentum and mass of $h_c(2P)$, $\epsilon$ is its polarization vector, and $a_i(i=1,2,3)$ are the radial wave functions. {  Notice that there is no $P\cdot q$ term in the wave function, because we have adopted the instantaneous approximation. Under this approximation, $P\cdot q = P\cdot {q}_{_{\bot}}$, so the radial wave function $a_i$ depends only on ${q}_{_{\bot}}^2$. The retardation effect in mesons was studied in Refs. \cite{retardation,retardation2}, and it was found that the retardation effect is very small for heavy mesons, indicating that the instantaneous approximation is a good approximation for heavy mesons. Moreover, this paper studies OZI-allowed strong decays, which occur near the threshold. The final-state particles move very slowly, and the contribution of relativistic corrections is relatively small. Therefore, the error introduced by using the instantaneous approximation is relatively small.}
	
	For the $0^{-}$ pseudoscalar $D^0$ meson, its wave function is \cite{kernel}
	\begin{equation}
		\varphi^{0^{-}}_{_{P_1}}(q_{_{1\bot}})=\left(\slashed{P}_{1}b_{1}+M_{1}b_{2}
		+\slashed{q}_{_{1\bot}}b_{3}+\frac{\slashed{P}_{1}\slashed{q}_{_{1\bot}}}{M_{1}}b_{4}\right)\gamma_{5},
	\end{equation}
	where $P_1$ and $M_1$ are the momentum and mass of $D^0$.
	For the $1^{-}$ vector $\bar{D}^{*0}$ with momentum $P_2$ and mass $M_2$, its wave function is \cite{wave1-}
	\begin{eqnarray}
		\varphi^{1^{-}}_{_{P_2}}(q_{_{2\bot}})=\epsilon_{2}\cdot q_{_{2\bot}}\left(c_{1}+c_{2}\frac{\slashed{P}_{2}}{M_{2}}+c_{3}\frac{\slashed{q}_{_{2\bot}}}{M_{2}}
		+c_{4}\frac{\slashed{P}_{2}\slashed{q}_{_{2\bot}}}{M_{2}^{2}}\right)
		+M_{2}\slashed{\epsilon}_{2}\left(c_{5}+c_{6}\frac{\slashed{P}_{2}}{M_{2}}\right)\nonumber\\
		+c_{7}\left(\slashed{q}_{_{2\bot}}\slashed{\epsilon}_{2}-\epsilon_{2}\cdot q_{_{2\bot}}\right)+c_{8}\left(\frac{\slashed{P}_{2}\slashed{\epsilon}_{2}\slashed{q}_{_{2\bot}}-\slashed{P}_{2}\epsilon_{2}\cdot q_{_{2\bot}}}{M_{2}}\right),
	\end{eqnarray}
	where $\epsilon_{2}$ is the polarization vector of $\bar{D}^{*0}$. 
	The radial wave function $b_{i}=b_{i}(-q^{2}_{1\bot})~(i=1,2,3,4)$ and $c_j=c_j(-q^2_{_{2\perp}})\ (j=1,...,8)$ are obtained numerically by solving the Salpeter equations corresponding to the pseudoscalar \cite{kernel} and vector mesons \cite{wave1-}, respectively.

	\section{Results and discussions}
	
	In the calculation, we chose the parameters as {  $\lambda=0.21$ GeV$^2$, $\alpha=0.06$ GeV, $\Lambda_{\text{QCD}}=0.27$ GeV, $m_d=0.311$ GeV, and $m_u=0.307$ GeV, and $m_c=1.75$ GeV \cite{1.75}. $m(D_{0})=1.865$ GeV, $m(D_{0}^{*})=2.007$ GeV, $m(D_{\pm})=1.87$ GeV, $m(D_{\pm}^{*})=2.01$ GeV \cite{PDG}.} Using the BS method, we obtained a total width of $69.6^{+40.4}_{-30.4}$ MeV for $h_{c}(2P)$, where the error is derived by arbitrarily varying all parameters by $\pm5$\%. When applying the $^3P_0$ method, we set the central value of its width to match that from the BS method, from which we deduced $\gamma=0.411^{+0.109}_{-0.133}$. 

Our results and the ones from other groups are listed in Table \ref{tab2} for comparison. As can be seen, our results are consistent with those of most theoretical studies. {  It can be seen that there are some differences between the results of this paper and those of Ref. \cite{wzh}. In Ref. \cite{wzh}, the charm quark mass is 1.62 GeV, whereas it is 1.75 GeV in this paper. However, the difference in the width mainly arises from the different values of $\gamma$, rather than from the charm quark mass. In this paper, $\gamma = 0.411$, while in Ref. \cite{wzh}, $\gamma = 0.35$; the square of their ratio is 1.38. The ratio of the width in this paper to that in Ref. \cite{wzh} is 1.45, which is close to 1.38, indicating that the difference in the width is mainly due to the difference in $\gamma$.}

{  We adopt $m_c = 1.75$ GeV, rather than $m_c = 1.62$ GeV as used in Ref. \cite{wzh}. This is because, in order to verify the correctness of the function ${\eta}_{_{P_{1}}}(q_{_{1\perp}})$ (and ${\eta}_{_{P_{2}}}(q_{_{2\perp}})$), we substituted it into the first expression of Eq. (\ref{sal2}) to obtain the wave function $\varphi^{++}$ and compared it with those obtained by directly solving the Salpeter equation. We found that for  $m_c = 1.62$ GeV, the $\varphi^{++}$ of the $D^*$
 in the small-$|\vec{q}|$ region was unsatisfactory, showing deviations. However, when $m_c = 1.75$ GeV, the results agreed very well. A possible reason for the deviation is that when solving for the $\varphi^{++}$ via ${\eta}_{_{P_{1}}}(q_{_{1\perp}})$,  
$M-\omega_1-\omega_2$ appears in the denominator. For $m_c = 1.62$ GeV and $m_d = 0.311$ GeV, 
$m_c + m_d$ is close to but slightly less than the $D^*$ mass $M$, making this denominator sensitive in the small-$|\vec{q}|$ region. Therefore, we take 
$m_c = 1.75$ GeV to avoid this sensitive region.}
	
	\begin{table}[ht]
		\centering \caption{Mass and decay width of $h_{c}(2P)$, and the adopted $\gamma$-value} \label{tab2}
		\setlength{\tabcolsep}{7pt} 
		\renewcommand{\arraystretch}{1} 
		\begin{tabular}{|c|c|c|c|c|c|c|c|}
			\hline
			& $^{3}P_{0}$ & BS & \cite{wzh} & \cite{wzh24}& \cite{wzh24}   & \cite{WZH26}    & \cite{gamaqu0.35}
			\\ \hline $\rm{Mass}~(\rm{MeV})$&3943&3943&3943&3940&3916&3934&3934
			\\ \hline $\Gamma(D^{0}\bar{D}^{\ast0}+{D^{\ast0}}\bar{D}^{0})$ & $34.2^{+7.6}_{-0.4}$ & $34.2^{+19.8}_{-15}$ & - & -& - & - & -
			\\\hline    $\Gamma(D^{+}\bar{D}^{\ast-}+{D^{\ast+}}\bar{D}^{-})$  &$35.4^{+8.2}_{-0.6}$&  $35.4^{+20.6}_{-15.4}$ & - & - & -& - & -
			\\\hline    $\Gamma_{\rm total}$ (MeV)    &$69.6^{+15.8}_{-1.0}$& $69.6^{+40.4}_{-30.4} $ &48 &64 &68&87  & 67
			\\ \hline $\gamma$&$0.411^{+0.109}_{-0.133}$&&0.35&0.234&0.217&0.4&0.35
			\\\hline
		\end{tabular}
	\end{table}
	
{ As can be seen from Table \ref{tab2}, the relative error of our approach using the BS method is much larger than that using the $^3P_0$ method. This is because the calculation within the BS method involves one more integral than that in the $^3P_0$ method. Comparing formulas in Eq. (\ref{amplitude3}) and Eq. (\ref{3P0am}), which are the amplitude formulas for the BS method and the $^3P_0$ method, respectively, the function $\eta$ in Eq. (\ref{amplitude3}) is an abbreviation for the integral in Eq. (\ref{right}). Therefore, the amplitude formula for the $^3P_0$ method contains only a single integral, while that for the BS method contains a double integral, leading to a large error in the width obtained by the latter. The parameter $\gamma$ is extracted as the ratio of the widths obtained from the BS method and the $^{3}P_{0}$ model. Since both methods employ the same interaction kernel and potential parameters, part of the uncertainty cancels in the ratio, leading to a relatively stable determination of $\gamma$. Therefore, the relative error of $\gamma$ is smaller than that of the width calculated using the BS method.}

\begin{figure}[htbp]
		\centering   
		\begin{subfigure}[b]{0.4\textwidth} 
			\centering      
			\includegraphics[width=\textwidth]{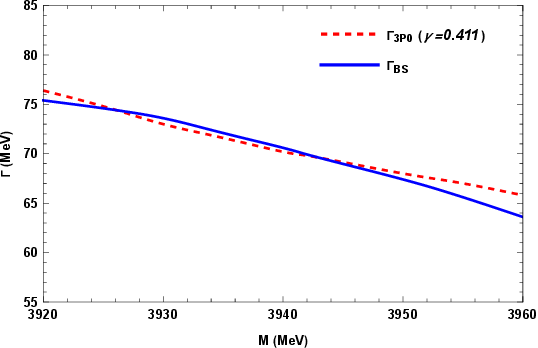}
			\caption{Width by the BS and $^3P_0$ methods} 
			\label{fig:sub1}    
		\end{subfigure}
		\hfill  
		\begin{subfigure}[b]{0.4\textwidth} 
			\centering
			\includegraphics[width=\textwidth]{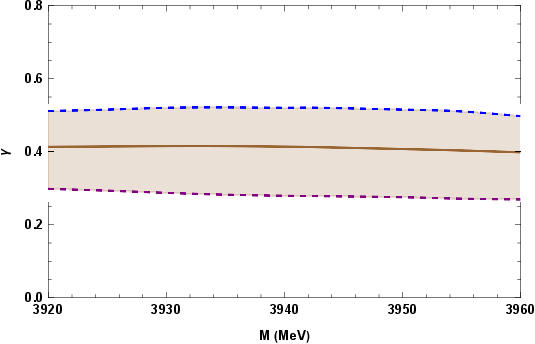}
			\caption{The predicted $\gamma$}  
			\label{fig:sub2}    
		\end{subfigure}
		\caption{Width and $\gamma$-value as functions of $h_{c}(2P)$-mass.}
		\label{fig2} 
	\end{figure}

	{  Currently, $h_{c}(2P)$ has not been observed experimentally, and its mass needs to be provided by theory. We previously predicted its mass to be 3943 MeV in Ref. \cite{Chang:2010kj}, where $m_c = 1.62$ GeV. If we adopt $m_c = 1.75$ GeV as in this paper, the $h_{c}(2P)$ mass becomes 3949 MeV, which is close to 3943 MeV. For comparison with Ref. \cite{wzh}, we adjust the free parameter $V_0$ in our calculation so that the $h_{c}(2P)$ mass is set to 3943 MeV. To eliminate this arbitrariness, it is necessary to present the dependence of the results on the mass of $h_{c}(2P)$.} In Figure \ref{fig2}, subfigure (a) presents the variation of the total width calculated using the BS method and the $^3P_0$ method with the mass of $h_{c}(2P)$, where $\gamma$ in the $^3P_0$ model is fixed at 0.411. Subfigure (b) shows the variation of the predicted $\gamma$-value with the  $h_{c}(2P)$ mass.	
	
{  From Figure \ref{fig2}(a), it can be seen that the change in the $h_{c}(2P)$ mass has a significant impact on the total width. However, as shown in Figure \ref{fig2}(b), this influence does not propagate to the $\gamma$ value. In other words, since gamma is the ratio of widths,  within the chosen parameter range, the effect of the mass change is almost canceled out in the ratio.}

{In summary, this paper presents a relativistic BS approach to OZI rule-allowed strong decays. A key feature of this formalism is that the effective coupling relevant to the $^{3}P_{0}$ model is not introduced as a free parameter, but is determined by the interaction kernel of the BS equation. This establishes a direct connection between the phenomenological $^{3}P_{0}$ coupling and the underlying quark–antiquark interaction, enabling a theoretical determination of $\gamma$ within the potential model from the same kernel that governs hadron spectra.}
	
	\vspace{0.7cm} {\bf Acknowledgments}
	
This work was supported by the National Natural Science Foundation of China (NSFC) under the Grant No. 12575097 and by the Natural Science Foundation of Guangxi Autonomous Region with Grant No. 2022GXNSFDA035068. T. Wang was supported by the NSFC under the Grant No. 12375085, and Z. Wang was supported by the NSFC under the Grant No. 12365013.

\end{document}